# Probing the bright exciton state in twisted bilayer graphene via resonant Raman scattering

*Matthew C. DeCapua,[1] Yueh-Chun Wu,[1] Takashi Taniguchi,[2] Kenji Watanabe[3] and Jun Yan[1,*]*

[1]Department of Physics, University of Massachusetts Amherst, Amherst, Massachusetts 01003, USA

[2] International Center for Materials Nanoarchitectonics, National Institute for Materials Science, Tsukuba, Ibaraki 305-0044, Japan

[3] Research Center for Functional Materials, National Institute for Materials Science, Tsukuba, Ibaraki 305-0044, Japan

[*]Corresponding Author: Jun Yan      **Email:** yan@physics.umass.edu

**Abstract:** The band structure of bilayer graphene is tunable by introducing a relative twist angle between the two layers, unlocking exotic phases, such as superconductor and Mott insulator, and providing a fertile ground for new physics. At intermediate twist angles around 10°, highly degenerate electronic transitions hybridize to form excitonic states, a quite unusual phenomenon in a metallic system. We probe the bright exciton mode using resonant Raman scattering measurements to track the evolution of the intensity of the graphene Raman G peak, corresponding to the $E_{2g}$ phonon. By cryogenically cooling the sample, we are able to resolve both the incoming and outgoing resonance in the G peak intensity evolution as a function of excitation energy, a prominent manifestation of the bright exciton serving as the intermediate state in the Raman process. For a sample with twist angle 8.6°, we report a weakly temperature dependent resonance broadening $\gamma \approx 0.07$ eV. In the limit of small inhomogeneous broadening, the observed $\gamma$ places a lower bound for the bright exciton scattering lifetime at 10 fs in the presence of charges and excitons excited by the light pulse for Raman measurement, limited by the rapid exciton-exciton and exciton-charge scattering in graphene.

The manipulation of interlayer interaction by controlling the relative lattice orientation between two stacked, atomically thin crystals has emerged as a powerful tool to engineer the electronic properties of two-dimensional (2D) material. Bilayer graphene is an outstanding example of such 'twistronic' materials[1–10]. In its preferred orientation, bilayer graphene has a Bernal stacking structure, with half the top layer lattice sites directly above the lattice sites of the bottom layer, and the other half centered within the hexagons of the bottom layer lattice[11–14]. Introducing a relative angle between the two sheets of graphene produces interference between the periodic potentials of the two layers, resulting in the formation of a moiré superlattice potential modulation which folds the Brillouin zone[15,16], and generates flat bands and van Hove singularities[17–21]. These flat bands emerge from interlayer interactions that lead to the hybridization of bands from individual layers, resulting in strong electron correlations and the development of a rich phase diagram, including superconducting[2,3,22], Mott insulating[1,2], and ferromagnetic states[4–6]. This has been demonstrated in bilayer graphene for small, "magic" twist angles[23]. Because of its interlayer origin, the relative twist angle between the two constituent layers of graphene has been identified as a crucial tunable parameter in the realization of these different phases. Understanding the twist angle's effect on the electronic interactions of graphene is thus critical to deepen our comprehension of graphene twistronics for future device applications.

For twist angles larger than the "magic" angle, the linear dispersion of graphene near the charge neutral point is preserved in the twisted bilayer band structure[21]. The group velocities of the conduction and valence bands are the same and thus run parallel to each other (see Fig. 1a). These parallel bands, formed when the conduction band Dirac cone of one layer is offset over the valence Dirac cone of the other layer by the twist angle, also contribute to an increase in the joint density of states at the resonance energy corresponding to the separation between the two bands. This resonance is tunable by adjusting the twist angle, going to higher energy for larger angles as the Dirac cones separate more in momentum space. Since there are two combinations to form parallel bands using the conduction and valence bands from each layer, two degenerate interlayer transitions result[7,8,12].

Recently it has been shown that interlayer excitonic states are possible in these parallel band regions, despite bilayer graphene being semi-metallic[8,24,25]. The two degenerate states, $X_{13}$ and $X_{24}$, rehybridize into an optically bright, symmetric combination $X_S$, and an optically dark, antisymmetric combination $X_A$[24]. The dark exciton is tightly bound and has a lifetime of about 1 ps[26]. The bright exciton is less spatially localized according to theoretical calculations, but has been shown to be capable of producing photoluminescence emission[26,27], an atypical phenomenon in a semi-metallic system. The lifetime of the bright exciton, although believed to be much shorter, so far lacks a reliable determination. The linewidth of resonance peaks in absorption spectra carries information on the lifetime, but it is in the meantime also complicated by inhomogeneous broadening[8,17,19].Transient absorption at the bright exciton energy after one- or two-photon excitation only reveals the dynamics of the dark exciton, as the optical response of the bright exciton is determined by the population and depletion of the longer lived dark excitons which serve as a bottleneck in the optical processes[25,26].

In this work, we use resonant Raman scattering to probe the optical properties of the twisted bilayer graphene bright exciton. The resonance condition can be met in two ways: incoming resonance, where the excitation energy matches the energy of the bright exciton state, and outgoing resonance, where the excitation is higher than the energy of the bright state by an amount equal to the energy of the graphene G phonon, 0.196 eV[28,29]. Previous measurements of the resonance profile have been performed either as a function of twist angle at fixed laser excitation energy, or as a function of laser energy at a fixed twist angle [7,9,18,30–32]. However, neither type of profiles showed a double resonance peak structure. Here, we report that by carefully measuring the resonance profile over a range of temperatures, we are able to resolve both the incoming and the outgoing peaks in graphene Raman intensity. Fitting the Raman intensity as a function of laser excitation energy, we determine the time over which the bright exciton provides significant contribution to the Raman cross section, setting a lower bound for the exciton coherence lifetime. Understanding these fundamental carrier dynamics is essential for future optoelectronic device applications using twisted graphene structures.

Our twisted bilayer graphene sample is fabricated with a tear-and-stack method[33], as shown in Fig.1b. We mechanically exfoliated graphene from bulk graphite onto silicon wafers with a 285 nm silicon oxide layer. Once a suitable piece of monolayer graphene was located (Fig.1b, left), we used a polypropylene carbonate (PPC) film mounted on a glass slide to pick up a flake of hexagonal boron nitride (hBN), roughly 80 nm in thickness, which we then used to pick up part of the monolayer graphene. This created a tear in the graphene as the part of the monolayer under the BN was lifted, while the exposed graphene remained on the chip (Fig.1b, middle). After lifting, we rotated the stage to the desired angle, and lowered the lifted graphene down, overlapping with the graphene on the chip (Fig. 1(b), right). Because the two layers originated from the same graphene crystal, presumably in the same orientation, this gave us deterministic control over the twist angle. The final step was to transfer the sample to a silicon chip with a 65 nm silicon oxide layer grown on top. This thickness was selected to create constructive interference and enhance the reflection contrast.[34]

Figure 2a shows differential reflectance and resonant Raman spectra of our sample at 3 Kelvin (K). The differential reflectance is obtained by subtracting and normalizing the twisted bilayer graphene reflection with respect to reflection from a nearby region of monolayer graphene covered by the same hBN flake. For our bilayer graphene with a twist angle of 8.6°, we found an increase in optical contrast for wavelengths around 688 nm (1.80 eV). We interpret this as an increase in absorption for photons in resonance with the bright exciton state. The correlation between the absorption peak position and the twist angle is consistent with previous studies in literature[7,15,30].

The Raman spectroscopy was performed using an 80 MHz white laser with a series of optical filters to select a particular wavelength. With a home-built grating system we achieve reasonably narrow laser linewidths of less than one nanometer. This allows us to select from a near continuum of excitation wavelengths at the expense of some peak broadening, as our source linewidth is still wider than typical Raman systems. We swept the laser excitation energy from 1.75 to 2.05 eV with a 5 meV resolution. Raman signal of the G band mode that has an energy of 196 meV is dispersed by a triple spectrometer and collected by a liquid nitrogen cooled CCD camera. We arrange the Raman peaks in Fig.2a according to the excitation photon energy. The intensities of the peaks were extracted from the area of a Lorentzian fit, and normalized by that of the silicon substrate Raman signature at 520 cm$^{-1}$ to correct for differences in power and spectral response of the optical components at different wavelengths. Due to power limitations and the reduction of Raman scattering intensity at longer wavelengths, for excitations above 600 nm the single layer graphene Raman signal was no longer visible above the noise, so only a linear background was deconvoluted from the bilayer response. As can be seen in the figure, the G mode intensity reaches a maximum with an excitation wavelength of 690 nm (1.79 eV) and then again at 625 nm (1.98 eV).

The first resonance at 1.79eV matches the bright exciton absorption feature in differential reflection spectrum, and we attribute it to the incoming resonance in the Raman process (Fig.2b left). We note that there exists some uncertainty in absorption measurements due to multi-reflection and interference that can cause oscillating artifacts in the reflection spectra of the multilayer structure. While our data at 3 K in Fig.2a is relatively clear, at higher temperatures we observe more complicated features that make interpretation of data more difficult, as shown in Supplementary Section 2. Nevertheless, the fact that we consistently see a dip at ~690nm at all temperatures, and that it corresponds nicely to the first Raman resonance support that the relevant feature is due to bright exciton absorption of the twisted bilayer graphene. The second resonance occurs at about 0.2 eV above the first resonance, matching well to the G band phonon energy. This supports its assignment as the outgoing Raman resonance: the incident photon at around 1.98eV does not resonate with the exciton state, but after creation of the G band phonon, the residual energy matches that of the bright exciton, leading to enhanced Raman cross section (Fig.2b right).

The detailed wavelength dependence of the graphene G peak intensity at T = 3 K as well as other temperatures are shown in Fig. 3a. At low temperatures, the incoming and outgoing resonances lead to an 'M' shaped resonance profile. As T increases, the dip between incoming and outgoing resonances flattens and becomes a plateau for temperatures above 200 K. We fit these normalized excitation profiles using a simplified time-dependent perturbation theory calculation,[15,30]

$$I(G) = \left| \frac{M}{(E - E_{X_S} - i\gamma)(E - E_{X_S} - \hbar\omega_G - i\gamma)} \right|^2 \tag{1},$$

where $M$ is the product of scattering matrix elements for electron-phonon and electron-photon interactions, $E$ is the energy of the laser excitation source, $E_{X_S}$ is the energy of the bright exciton state, $\hbar\omega_G$ is the energy of the G phonon, and $i\gamma$ is a broadening term related to the lifetime of the scattering process. Previous measurements have shown that the G-band energy has only a very weak temperature dependence[35], consistent with our observation here. To simplify the fit, we assume temperature independent $\hbar\omega_G$. At each temperature, the resonance profile can be fit using Equation (1) to determine the three parameters ($M$, $E_{Xs}$, and $\gamma$). Figure 3 b-d show these fitting parameters as the scattered black squares. The temperature dependence of these parameters is roughly linear, and can be fitted by the following linear functions (red lines in Fig.3 b-d):

$$\gamma(T) = (5.1 \pm 3.3) \times 10^{-6} \, T + (0.071 \pm 0.005) \tag{2},$$

$$E_{Xs}(T) = (-1.7 \pm 2.0) \times 10^{-6} \, T + (1.793 \pm 0.003) \tag{3},$$

$$M(T) = (2.7 \pm 0.8) \times 10^{-6} \, T + (0.017 \pm 0.001) \tag{4}.$$

In Eqs. (2-4), the parameters $E_{Xs}$ and $\gamma$ are expressed in eV, $M$ in eV$^2$ and the temperature $T$ in Kelvin. Using Eqs. (1-4), we then perform a global fit to our experimental data in Fig.3a (colored lines).

The temperature dependence of $M$, $E_{Xs}$, and $\gamma$ is weak, with the uncertainty of the fitted slope larger than the slope itself for the resonant energy $E_{Xs}$, and significant error bars of roughly 60% of the determined value for the slope of $\gamma(T)$. Nonetheless, the good match between experiment and fit indicates that Eq. (1), together with the temperature evolution of the fitting parameters, provides adequate description of the observed resonance profile.

The 'M' shaped resonance profile observed in Fig.3 is a manifestation of the dominant role of twisted bilayer graphene bright exciton in the Raman process near the incoming and outgoing resonances. Assuming that the bright exciton is a reasonably well-defined resonance state, the measured broadening $\gamma$ of 0.07 eV at low temperatures that increases up to 0.09 eV at room temperature provides interesting insight into the fundamental properties of the bright exciton. This $\gamma$ is much larger than the linewidth of our laser and the G band phonon. When the laser excitation is off-resonance, i.e. the difference between E and $E_{Xs}$, $E_{Xs} + \hbar\omega_G$ is much larger than $\gamma$, the Raman process is fast, almost instantaneous. When on-resonance however, either incoming or outgoing, the Raman process becomes slower due to lifetime effects of the bright exciton. Raman is a coherent process and the time delay between absorption of incident photon and emission of scattered photon is determined by the fastest dynamic process that scatters the exciton and causes loss of coherence. Our measured $\gamma$ corresponds to a scattering lifetime on the order of 10 femtoseconds. Previous reports of this quantity range from 0.1 eV to 0.25 eV (~6 to ~2 fs) for similar twist angles[8,24,30,36]. It is also of interest to note that ultrafast dynamics study of graphene revealed that electron-electron hot carrier scattering is also of the order of 20 fs[37], comparable to what we observe in the resonant Raman measurement here. The bright exciton $X_s$ state is weakly bound[22,25,38], and extends much more spatially than the dark exciton. The bright exciton can thus quickly scatter off another exciton, or charge, or disorder, lose its coherence and can no longer contribute to the Raman signal. Note however, that these weakly bound excitons near the K points of twisted bilayer graphene are still longer lived than the M point excitons found in single layer graphene[39]. Under illumination of laser excitation for our Raman measurement, we estimate the dark exciton number density to be of order $10^{11}$ cm$^{-2}$. This is low compared to typical ultrafast pump-probe measurements. Further studies at higher excitation level may provide interesting insights regarding density dependence of the broadening and the coherence lifetime of the bright excitons.

To conclude, we performed reflection contrast spectroscopy and tunable excitation Raman spectroscopy on bilayer graphene with a twist angle of 8.6°. The two measurements demonstrate consistent resonance effects on the same device due to the symmetric bright exciton arising from interactions between the two graphene layers. Using a tunable laser excitation source with sub-nanometer resolution, we sweep through the resonant energy and generate a resonance Raman profile that reveal both the incoming and outgoing resonant Raman when the sample is cooled cryogenically. The weak temperature dependence of the bright exciton energy $E_{Xs}$ is in contrast to bright excitons in transition metal dichalcogenide systems, which exhibit significant redshift in energy with increasing temperature due to thermal expansion modifying the band structure and exciton-phonon interactions [40–43]. The weakly temperature dependent scattering lifetime $\gamma$ of the bright excitonic state is also determined quantitatively to be about 10 fs. We find that previous studies may have slightly overestimated this broadening.

**Supplementary Material:** determination of twist angle from the R' Raman peak activated by moiré potential; temperature dependence of reflection spectra.

**Acknowledgment:**

This work is supported by National Science Foundation (NSF, Grant number: DMR-2004474). K.W. and T.T. acknowledge support from the Elemental Strategy Initiative conducted by the Japan Ministry of Education, Culture, Sports, Science and Technology (MEXT, Grant Number: JPMXP0112101001), and the Japan Society for the Promotion of Science Grants-in-Aid for Scientific Research (JSPSKAKENHI Grant Numbers JP20H00354) and the Japan Science and Technology Agency Core Research for Evolutional Science and Technology (CREST, Grant: JPMJCR15F3).

**DATA AVAILABILITY**

The data that supports the findings of this study are available within the article and its supplementary material.

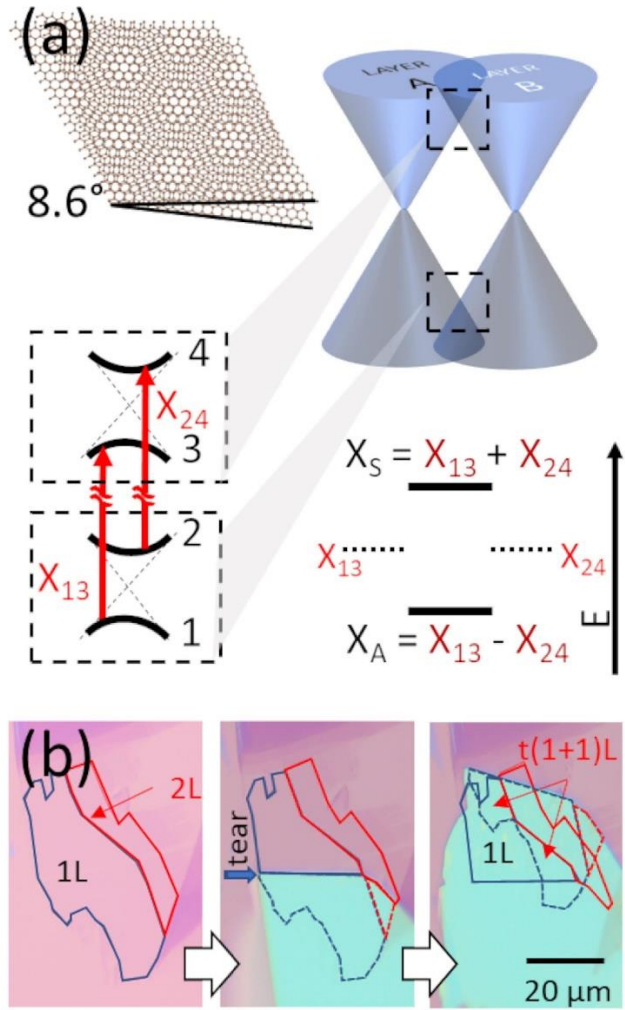

**FIG. 1.** (a) Schematic showing the formation of moiré patterns by introducing a relative angle of 8.6° between two graphene layers, as well as the band structure near the Fermi-level. The avoided crossing where the Dirac cones overlap are magnified in the dashed insets. The degenerate electronic transitions $X_{13}$ and $X_{24}$ are hybridized into symmetric and antisymmetric states. (b) Progression showing the tear-and-stack method to produce twisted bilayer graphene.

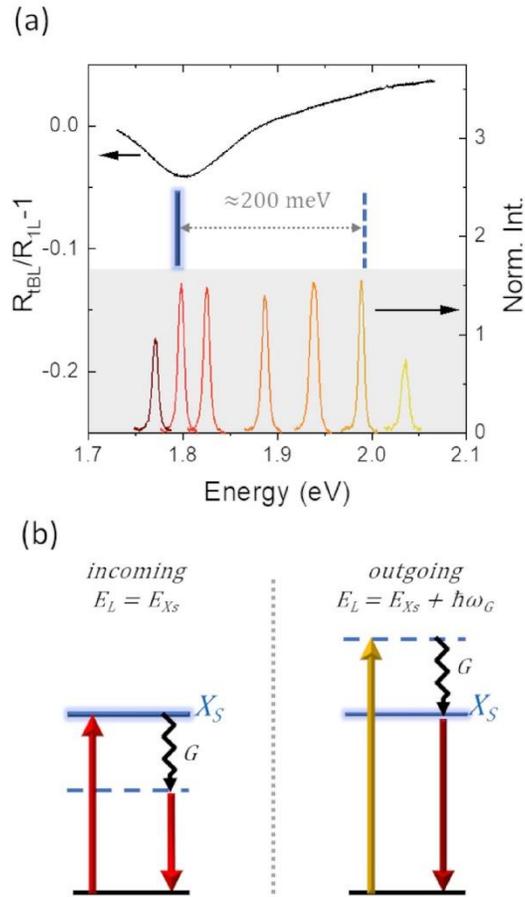

**FIG. 2.** (a) Top: Differential reflection of the twisted bilayer graphene relative to single layer graphene at 3K. Bottom: Graphene Raman G peak profile measured at 3K for a range of excitation energies. The Raman peaks are centered at the energy of the laser excitation. The scale (right axis) is the intensity normalized by the intensity of the silicon Raman peak at 520 cm$^{-1}$. (b) Sketch of the incoming and outgoing resonances mediated by the $E_{2g}$ graphene G-band phonon.

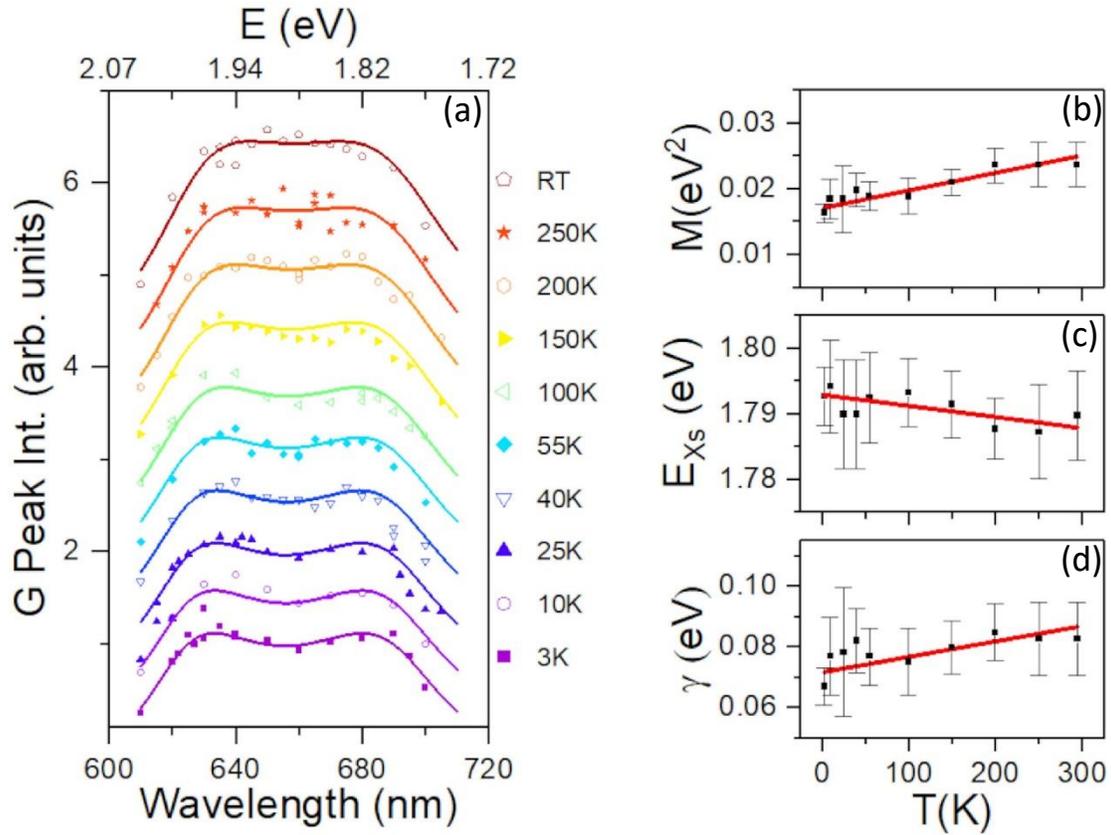

**FIG. 3.** (a) Plot of graphene Raman G peak intensity dependence on excitation energy. The intensity is normalized by the intensity of the silicon Raman peak at 520 cm$^{-1}$ and shifted for clarity. Experimental data (symbols) is fit to a simplified time-dependent perturbation model using $\gamma$, E$_{Xs}$, and scattering matrix element M as fitting parameters. (b)-(d) The fitting parameters (symbols) and their linear fits (red line).

# Supplementary Material: Probing the bright exciton state in twisted bilayer graphene via resonant Raman scattering

*Matthew C. DeCapua,[1] Yueh-Chun Wu,[1] Takashi Taniguchi,[2] Kenji Watanabe[3] and Jun Yan[1,*]*

[1]Department of Physics, University of Massachusetts Amherst, Amherst, Massachusetts 01003,

USA

[2] International Center for Materials Nanoarchitectonics, National Institute for Materials Science,

Tsukuba, Ibaraki 305-0044, Japan

[3]Research Center for Functional Materials, National Institute for Materials Science, Tsukuba,

Ibaraki 305-0044, Japan

## 1. Determination of twist angle

The relative angle between the two graphene layers introduced using the tear-and-stack method was ascertained both by inspection of the optical image and by spectroscopic features, including the position of the R' Raman peak. This peak arises from an intravalley double-resonant process involving scattering off the superlattice potential, which transfers momentum that is dependent on the twist angle. Therefore, the peak position will shift for different angles.[1] Using 532 nm wavelength laser excitation, we observe the R' peak at 1619 cm$^{-1}$, in good agreement with the angle 8.6° we obtain from the optical images, as well as the incoming resonance energy.[2,3]

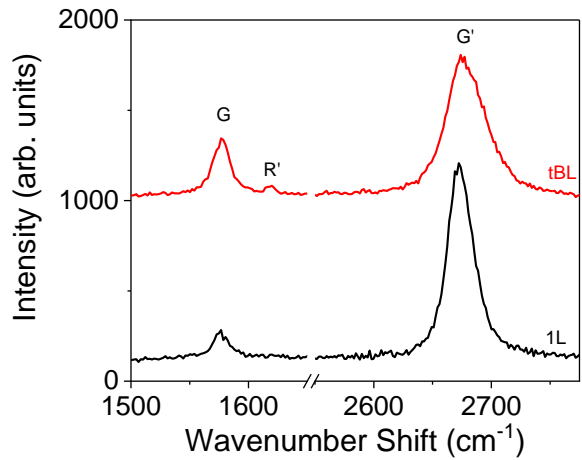

**SFIG.1**. Raman spectra of twisted bilayer (tBL) and single layer (1L) graphene measured by 532 nm cw laser excitation. The energy of the R′ peak is in good agreement with expectations from the 8.6° twist angle.

## 2. Determination of bright exciton energy using reflection contrast

As shown in the main text, the twisted bilayer graphene shows a dip in reflection relative to monolayer graphene, which we attribute to an increase in absorption of photons in resonance with the bright exciton state. Though sensitive to experimental conditions, the dip is reproducible from 3K to room temperature. In SFig.2, the y-axis is the intensity of the reflection from the twisted bilayer region which exhibited resonance, measured in counts, subtracted and divided by the intensity of the reflection of a nearby monolayer region. The plots are shifted for clarity. Also shown is the linewidth obtained for the base temperature measurement by fitting to a Lorentzian. The arrows are added for clarity and are not to scale. The linewidth (~160 meV) of the feature is consistent with previously observed absorption spectra[4,5]. The additional broadening relative to the $\gamma$ (~70 meV) extracted from the Resonant Raman Profile in the main text can be attributed to radiative and nonradiative decay channels that do not participate in the Raman scattering.

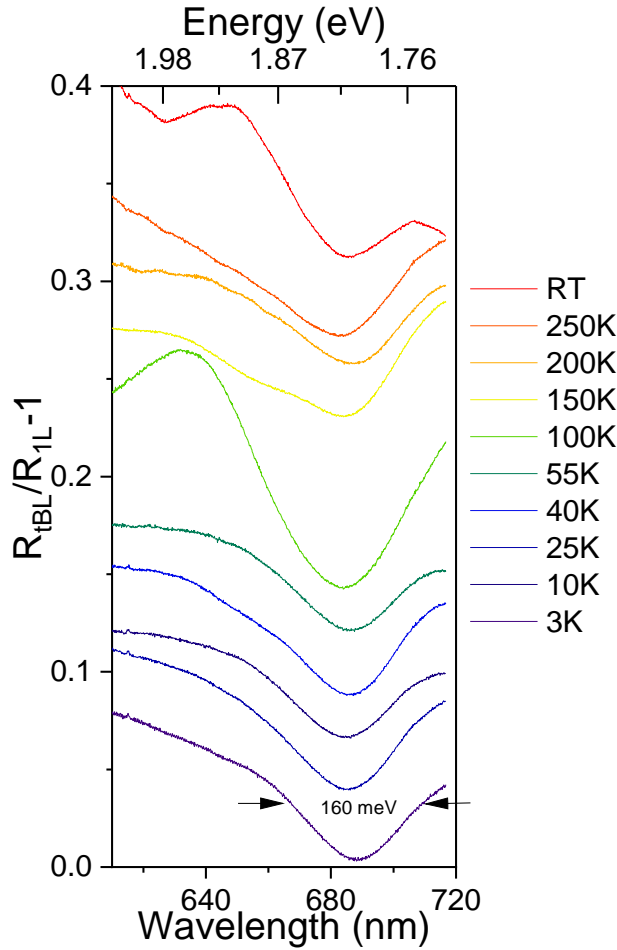

**SFIG.2**. Temperature dependence of the normalized reflection spectra. The linewidth of the bright exciton absorption feature is about 160 meV.